\documentclass[twocolumn,showpacs,preprintnumbers,amsmath,epsfig,amssymb,aps]{revtex4}

\usepackage{graphicx}
\usepackage{dcolumn}
\usepackage{bm}

\begin{document}

\newcommand{\be}{\begin{eqnarray}}
\newcommand{\ee}{\end{eqnarray}}
\newcommand{\gein}{$G_{EI}$ }
\newcommand{\gen}{$G_E$ }

\title{Percolation in living neural networks}

\author{Ilan Breskin\footnote[1]{ These authors contributed equally to this work.}}
\author{Jordi Soriano\footnotemark[1]}
\author{Elisha Moses\footnote[2]{Electronic address: {\tt elisha.moses@weizmann.ac.il}.}}
\author{Tsvi Tlusty}
\affiliation{Department of Physics of Complex Systems. Weizmann
Institute of Science. Rehovot 76100, Israel.}

\date{\today}

\begin{abstract}

We study living neural networks by measuring the neurons' response to a global electrical
stimulation. Neural connectivity is lowered by reducing the synaptic strength, chemically
blocking neurotransmitter receptors. We use a graph--theoretic approach to show that the
connectivity undergoes a percolation transition. This occurs as the giant component
disintegrates, characterized by a power law with critical exponent $\beta \simeq 0.65$.
$\beta$ is independent of the balance between excitatory and inhibitory neurons and
indicates that the degree distribution is gaussian rather than scale free.

\end{abstract}



\maketitle

Representing complex structures as connected graphs yields a simplification that retains
much of their functionality. It is therefore natural that network connectivity emerges as
the fundamental feature determining statistical properties, including the existence of
power laws, clustering coefficients and the small world phenomenon \cite{Barabasi-2002}.
While experimental access to  man--made networks such as the WWW or e--mail is feasible
\cite{EckmannMoses-2002}, biological ones such as the genetic networks must be
painstakingly revealed node by node \cite{HybridProtein-2005}. The connectivity in living
neural networks is even more difficult to uncover \cite{White-1986}, since connections
are hard to identify \cite{Markram-2003,Chan-2004} and typically differ from brain to
brain and from culture to culture \cite{Marom-2002,Jimbo-1999}. Unraveling the neural
wiring diagram even in small cultures, with $\sim 10^5$ neurons and $\sim 10^7$
connections, is presently not feasible, although some progress has been attained in the
study of the link between neural connectivity and information coding
\cite{Chan-2004,Segev-2004,Ofer-2006}.

Neural cultures derived from rat hippocampus develop into networks that display bursts of
activity, governed by the presence of both excitatory and inhibitory neurons
\cite{Maeda-1995,Marom-2002}. In this Letter we present a novel experimental approach to
quantify statistical properties of such networks, and study them in terms of percolation
on random graphs \cite{Stauffer,Newman-2001}. We control the connectivity of the entire
network, gradually reducing the synaptic strength by means of chemical application. The
initially connected network progressively breaks down into smaller clusters until a fully
disconnected network is reached. The weakening of the network is quantified and the
distribution of sizes of connected components determined by analyzing the neurons'
response to stimulations applied simultaneously to the entire network. Viewed inversely,
as the network's connectivity increases, a percolation transition occurs at a critical
synaptic strength with the emergence of a giant component, which increases as a power law
with an exponent $\beta \simeq 0.65$.

\begin{figure}[!hb]
\begin{center}
\includegraphics[width=7.6cm]{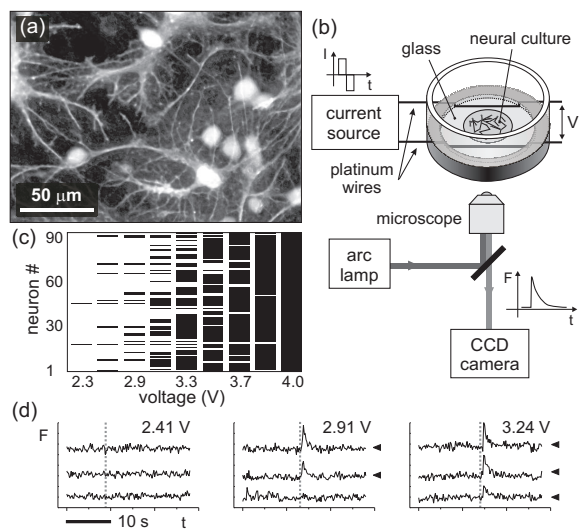} \vspace{-0.2cm}
\caption{(a) Fluorescence image of a small region of the neural culture. Bright spots are
cell bodies. Neural connections, mostly dendrites, are visible. (b) Sketch of the
experimental setup. (c) Activity plot of the neural response. Black lines indicate those
neurons that respond to the excitation. (d) $F(t)$ signal for 3 neurons at increasing
voltages. Vertical lines show the excitation time, and arrows the responding neurons.}
\label{Fig:setup}
\end{center}
\end{figure}

{\it Experimental setup and procedure.--} Experiments were performed on primary cultures
of rat hippocampal neurons, plated on glass coverslips following the procedure described
by Papa {\it et al.} \cite{Papa-1995} (Fig.\ \ref{Fig:setup}(a)). The cultures were used
14--20 days after plating. The neural network includes $N \simeq 2.5 \times 10^5$
neurons. The neural culture is placed in a chamber that contains two parallel platinum
wires fixed at the bottom and separated by $15$ mm (Fig.\ \ref{Fig:setup}(b)). The
neurons are electrically stimulated by applying a $20$ msec bipolar pulse through the
wires. The current is controlled and gradually increased between subsequent pulses, while
the corresponding voltage drop $V$ is measured with an oscilloscope. The chamber is
mounted on a Zeiss inverted microscope with a $10$X objective, and neuronal activity is
monitored using the fluorescent calcium indicator Fluo-4. Images were captured with a
cooled charge--coupled device (CCD) camera at a rate of $5$ frames/sec, and processed to
record the fluorescence intensity $F$ of a sample of the culture including $n \simeq 400$
individual neurons as a function of time (Fig.\ \ref{Fig:setup}(d)). The images and the
fluorescent signal are further analyzed to reject glia cells \cite{Supplementary}. Neural
spiking activity is detected as a sharp increase of the fluorescence intensity
\cite{Supplementary}.

The connectivity of the network was gradually weakened by adding increasing amounts of
CNQX (6-cyano-7-nitroquinoxaline-2,3-dione), the antagonist of the AMPA
(alpha-amino-3-hydroxy-5-methyl-4-isoxazolepropionic acid) type receptors in the
glutamate synapses of excitatory neurons. NMDA (N-methyl-D-aspartate) receptors were
completely blocked with $20$ $\mu$M of the corresponding antagonist APV
(2-amino-5-phosphonovalerate), enabling the study of network breakdown due solely to
CNQX. To study the role of inhibition, we performed experiments with inhibitory neurons
either active or blocked with $40$ $\mu$M of the GABA (Gamma-aminobutyric acid) receptor
antagonist bicuculine. From here on, we label the network containing both excitatory and
inhibitory neurons by $G_{EI}$, and the network with excitatory neurons only by $G_E$.

The response of the network for a given CNQX concentration was measured as the fraction
of neurons $\Phi$ that fired in response to the electric stimulation at voltage $V$
(Fig.\ \ref{Fig:setup}(c)). Response curves $\Phi(V)$ were obtained by increasing the
stimulation voltage from $2$ to $6$ V in steps of $0.1 - 0.5$ V. Between 6 and 10
response curves were measured per experiment, each at a different CNQX concentration.
Measurements were completed within 4 h, at the end of which the culture was washed of
CNQX to verify that the initial network connectivity was recovered.

{\it Model.--} To elucidate the relation between the topology of the living neural
network and the observed neural response, we consider a simplified model of the network
in terms of bond--percolation on a graph. The neural network is represented by the
directed graph $G$. Our main simplifying assumption is the following: \ A neuron has a
probability $f=f(V)$ to fire as a direct response to the externally applied electrical
stimulus, and it always fires if any one of its input neurons fire. This ignores the fact
that more than one input is needed to excite a neuron, and that connections are gradually
weakened rather than abruptly removed. The model also ignores the presence of feedback
loops and recurrent activity in the neural culture. However, we verified with numerical
simulations that relaxing these assumptions does not affect the validity of the model
\cite{Supplementary}.

Evidently, the firing probability $\Phi (f)$ increases with the connectivity of $G$,
because any neuron along a directed path of inputs may fire and excite all the neurons
downstream. All the upstream neurons that can thus excite a certain neuron define its
input--cluster or excitation--basin. It is therefore convenient to express the firing
probability as the sum over the probabilities $p_{s}$ of a neuron to have an input
cluster of size $s-1$,
\begin{eqnarray}
\Phi (f) &\nonumber=&f+(1-f)P\left( \text{any input neuron fires}\right)\\
&\nonumber=&f+(1-f)\sum_{s=1}^{\infty }p_{s}\left( 1-\left( 1-f\right)
^{s-1}\right) \\
&\label{Eq:Phi}=&1-\sum_{s=1}^{\infty }p_{s}\left( 1-f\right) ^{s},
 \end{eqnarray}%
where we used the probability conservation $\sum\nolimits_{s}p_{s}=1$. It is readily seen
that $\Phi (f)$ increases monotonically with $f$ and ranges between $\Phi (0)=0$ and
$\Phi (1)=1$. The deviation of  $\Phi (f)$ from linearity manifests the connectivity of
the network (for disconnected neurons $\Phi (f)=f$). Eq.\ (\ref{Eq:Phi}) indicates that
the observed firing probability $\Phi (f)$ is actually one minus the generating function
$H(x)$ (or the $z$--transform) of the cluster--size probability $p_{s}$
\cite{Shante-1971}, $H(x)=\sum_{s=1}^{\infty }p_{s}x^{s}=1-\Phi (f)$, where $x=1-f$. One
can extract from $H(x)$ the input--cluster size probabilities $p_{s}$, formally by the
inverse $z$--transform, or more practically, in the experiment, by fitting $H(x)$ to a
polynomial in $x$.

Once a giant component emerges the observed firing pattern is significantly altered. In
an infinite network, the giant component always fires no matter what the firing
probability $f$ is. This is because even a very small $f$ suffices to excite one of the
infinitely many neurons that belong to the giant component. We account for this effect by
splitting the neuron population into a fraction $g$ that belongs to the giant component
and always fires and the remaining fraction $1-g$ that belongs to finite clusters:
\begin{eqnarray}
\Phi (f) &\nonumber=&g+(1-g)\left[f+(1-f)P\left( \text{any inp. neur. fires}\right)\right]\\
&=&1-(1-g)\sum_{s=1}^{\infty }p_{s}\left( 1-f\right) ^{s}.
 \end{eqnarray}%
As expected, at the limit of almost no self--excitation $f\rightarrow 0$ only the giant
component fires, $\Phi (0)=g$, and $\Phi (f)$ monotonically increases to $\Phi (1)=1$.
With a giant component present the relation between $H(x)$ and the firing probability
changes, obtaining
\begin{equation}\label{Eq:Hx} H(x)=\sum_{s=1}^{\infty }p_{s}x^{s}=1-\Phi (f).
\end{equation} In reality, the giant component is not infinite and it is measured from
a sample which has $n$ neurons. Therefore, it fires only after a non--zero, though small,
firing probability $f_{T}$ is exceeded. To estimate this finite size threshold we note
that when we measure a giant component of size $gn$, then the firing probability is
\begin{equation}
\Phi (f)\simeq g\left( 1-\left( 1-f\right) ^{ng}\right) \simeq g\left( 1-e^{-fgn}\right).
\end{equation}
This probability becomes significant at the threshold
\begin{equation}\label{Eq:finite-size}
f_{T}\simeq (gn)^{-1}.
\end{equation}

\begin{figure}[!ht]
\begin{center}
\includegraphics[width=6.5cm]{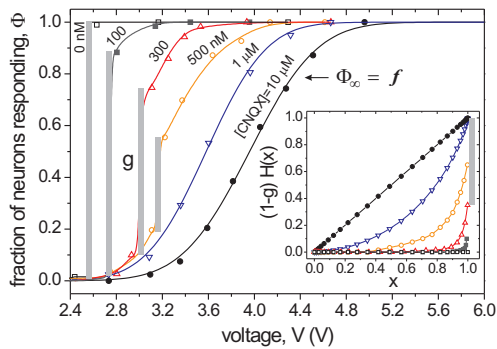} \vspace{-0.2cm}
\caption{Response curves $\Phi(V)$ for 6 concentrations of CNQX and $n=450$ neurons. The
grey bars show the size of the giant component. Lines are a guide to the eye except for
$1$ $\mu$M and $10$ $\mu$M that are fits to error functions, with $V_0 = 3.96$ V and
$\sigma_0=0.46$ V for $10$ $\mu$M. Inset: Corresponding $H(x)$ functions. The bar shows
the size of the giant component for $300$ nM.} \label{Fig:ErfCurves}
\end{center}
\end{figure}

{\it Measured network response.--} An example of the response curves $\Phi(V)$ for a
\gein network with $n = 450$ neurons measured at $6$ different concentrations of CNQX is
shown in Fig.\ \ref{Fig:ErfCurves}. At one extreme, with [CNQX] $=0$ the network is fully
connected, and a few neurons with low firing threshold suffice to activate the entire
culture. This leads to a very sharp response curve that approaches a step function, where
all neurons form a single cluster that comprises the entire network. At the other
extreme, with high concentrations of CNQX ($\simeq$ 10 $\mu$M) the network is completely
disconnected, the response curve rises moderately, and is given by the individual
neurons' response. $\Phi(V)$ for individual neurons (denoted as $\Phi_{\infty}(V)$) is
well described by an error function $\Phi(V)=0.5+0.5\;
\mathrm{erf}\left(\frac{V-V_0}{\sqrt 2 \,\sigma_0}\right)$. This indicates that the
firing threshold of a neuron in the network follows a gaussian distribution with mean
$V_0$ and width $2\sigma_0$.

Intermediate CNQX concentrations induce partial blocking of the synapses. As the network
breaks up neurons receive on average fewer inputs and a stronger excitation has to be
applied to light up the entire network. The response curves are gradually shifted to
higher voltages as [CNQX] increases. Initially, some neurons break off into separated
clusters, while a giant cluster still contains most of the remaining neurons. The
response curves are then characterized by a big jump that corresponds to the biggest
cluster ({\it giant component}), and two tails that correspond to smaller clusters of
neurons with low or high firing threshold (Fig.\ \ref{Fig:ErfCurves}). Error functions
describe these tails well. Beyond these concentrations ([CNQX] $\gtrsim 500$ nM for \gein
networks) a giant component cannot be identified and the whole response curve is then
also well described by an error function.

{\it Giant component.--} The biggest cluster in the network characterizes the giant
component, $g$. Experimentally, it is measured as the biggest fraction of neurons that
fire together in response to the electric excitation. For each response curve, $g$ is
measured as the biggest jump $\Delta \Phi$, as shown by the grey bars in Fig.\
\ref{Fig:ErfCurves}. The size of the giant component is considered to be zero when a
characteristic jump cannot be identified, or when the jump is comparable to the noise of
the measurement, which is typically about 4\% of the number of neurons measured.

We studied the size of the giant component in a range of CNQX concentrations spanning
almost three orders of magnitude from $0$ nM to $10$ $\mu$M in logarithmic scale. We
define the control parameter $c= 1/(1+[\mathrm{CNQX}]/K_d)$  as a measure of the synaptic
strength, where the dissociation constant $K_d = 300$ nM is the concentration of CNQX at
which $50$\% of the receptors are blocked \cite{Honore-1988}. The parameter $c$
characterizes the connectivity probability between two neurons, and takes values between
$0$ (full blocking, independent neurons) and $1$ (full connectivity). Conceptually, it
quantifies the number of receptor molecules that are not bound by the antagonist CNQX and
therefore are free to activate the synapse.

\begin{figure}[!ht]
\begin{center}
\includegraphics[width=6.4cm]{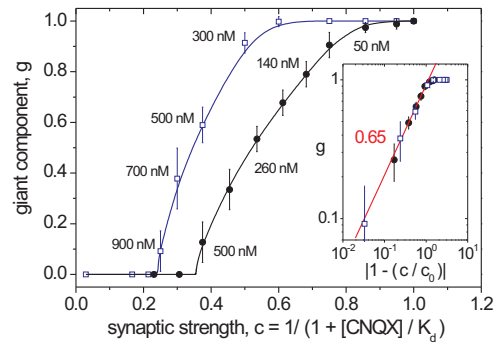} \vspace{-0.2cm}
\caption{Size of the giant component as a function of the synaptic strength $c$, averaged
over $18$ \gein networks ($\bullet$), and $6$ \gen networks ($\square$). Lines are a
guide to the eye. Some CNQX concentrations are indicated for clarity. Inset: Log--log
plot of the power law fits $g \sim |1-c/c_o|^{\beta}$. The slope $0.65$ corresponds to
the average value of $\beta$ for the two networks.} \label{Fig:giant}
\end{center}
\end{figure}

Fig.\ \ref{Fig:giant} shows the breakdown of the network for both \gein and \gen
networks. The giant component for \gein networks breaks down at much lower CNQX
concentrations compared with \gen networks, and one can think of the effect of inhibition
on the network as effectively reducing the number of inputs that a neuron receives on
average. For \gein networks, the giant component vanishes at [CNQX] $\simeq 600$ nM,
while for \gen networks the critical concentration is around $1000$ nM.

The behavior of the giant component indicates that the neural network undergoes a
percolation transition, from a set of small, disconnected clusters of neurons to a giant
cluster that contains most of the neurons. At the vicinity of the emergence of the giant
component, the percolation transition is described by the power law $g \sim |1 -
c/c_o|^{\beta}$. Power law fits for \gein and \gen networks give the same $\beta$ within
the experimental error (inset of Fig.\ \ref{Fig:giant}), with $c_o = 0.36 \pm 0.02$,
$\beta = 0.66 \pm 0.05$ for \gein, and $c_o = 0.24 \pm 0.02$, $\beta = 0.63 \pm 0.05$ for
\gen.

{\it Cluster distribution analysis.--} The construction of the experimental function
$H(x)$ defined in Eq.\ (\ref{Eq:Hx}) allows the fit of a polynomial
$\sum\nolimits_{s}p_{s}x^s$ to determine the size distribution $p_s(s)$ for clusters that
do not belong to the giant component. Since $f \equiv \Phi_{\infty}(V)$ is the response
curve for individual neurons (Fig.\ \ref{Fig:ErfCurves}) and $x=1-f$, the function $H(x)$
for each response curve is obtained by plotting $1-\Phi(V)$ as a function of
$1-\Phi_{\infty}(V)$. For curves with a giant component present, its contribution is
eliminated and the resulting curve normalized by the factor $1-g$.

\begin{figure}[!ht]
\begin{center}
\includegraphics[width=7.4cm]{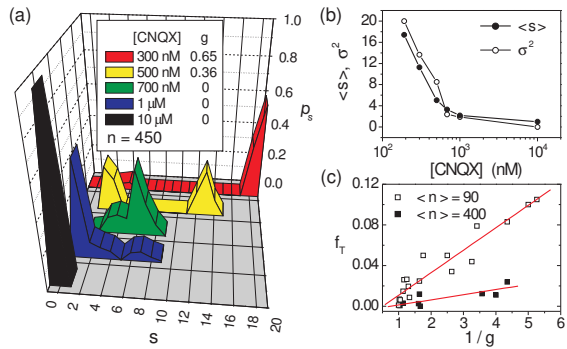} \vspace{-0.2cm}
\caption{(a) Cluster size distribution $p_s(s)$ for the experiment shown in Fig.\
\ref{Fig:ErfCurves}. (b) Average cluster size, $\langle s \rangle$, and variance of the
cluster size distribution, $\sigma^2 = \langle s^2 \rangle - \langle s \rangle ^2$, as a
function of the concentration of CNQX, averaged over $15$ \gein networks. (c) Giant
component firing threshold $f_T$ as a function of $1/g$ for two groups of experiments
with $90$ ($\square$) and $400$ ($\blacksquare$) neurons measured, and for CNQX
concentrations between $0$ and $500$ nM. Lines are least square fits.} \label{Fig:Ps}
\end{center}
\end{figure}

The inset of Fig.\ \ref{Fig:ErfCurves} shows $H(x)$ for the response curves shown in the
same figure. The corresponding $p_s(s)$ distributions, shown in Fig.\ \ref{Fig:Ps}(a),
are obtained from polynomial fits up to order $20$ \cite{Supplementary}. Since the
cluster analysis is sensitive to experimental resolution, it is limited to $g \lesssim
0.8$, which corresponds to [CNQX] $\gtrsim 200$ nM for \gein networks. Experimental
resolution also limits the observation of very small clusters for [CNQX] $\lesssim 500$
nM, since they are associated with values of $\Phi(V)$ close to 1 \cite{Supplementary}.
Hence, the cluster size distribution of Fig.\ \ref{Fig:Ps}(a) shows the correct behavior,
but not the precise details. Overall, as shown in Fig.\ \ref{Fig:Ps}(b), the clusters
start out relatively big and with a broad distribution in sizes, to rapidly become
smaller with a narrow distribution for gradually higher concentrations of CNQX. Isolated
peaks in $p_s(s)$ indicate non tree--like clusters outside the giant component, in
contrast to the model. This hints at the persistence of loops and at a strong local
connectivity. While our sample covers only part of the culture, it does represent the
statistics of the whole population. Sample size affects the noise level, but the overall
cluster size distribution of Fig.\ \ref{Fig:Ps}(a) is representative of the whole
network. Our assumption that one input suffices to excite a neuron leads to an
under--estimation of the cluster sizes \cite{Supplementary}, probably in direct
proportion to the number of inputs needed to excite a neuron, which is on the order of
ten \cite{Ofer-2006}.

Finite size effects are observed in the behavior of the firing threshold for the giant
component $f_T$, which increases linearly with $1/g$, as predicted in Eq.\
(\ref{Eq:finite-size}). $f_T(g)$ is measured for each concentration as the value of $f$
at which the giant component fires. Fig.\ \ref{Fig:Ps}(c) shows the results for two
groups of experiments with $\langle n \rangle = 90$ and $\langle n \rangle = 400$ neurons
measured. Linear fits provide slopes $\simeq 0.02$ and $\simeq 0.005$, which are of the
same order of magnitude as $1/n$, with $n$ the number of neurons measured.

{\it Discussion.--} Neural cultures turn out to be an experimental systems in which a
clear percolation transition can be mapped out in detail. The graph approach has proven
remarkably successful in supplying a simplified picture for a highly complex neuronal
culture, yielding quantitative measures of the connectivity. The measured exponent
$\beta$ appears to be independent of the culture details, such as the ratio between
excitation and inhibition or the variance between different cultures. Since $\beta$
characterizes the distribution of connections per node, it is possible to estimate the
connectivity distribution by comparing the experimental value of $\beta$ with the one
obtained from simulations or theoretical developments.

Numerical simulations of our model with a gaussian degree distribution provide $\beta
=0.66 \pm 0.05$, as in the experiments \cite{Supplementary}. Percolation on directed
random graphs with power law degree distribution $p_k(k) \sim k^{-\lambda}$ gives $\beta$
equal or larger than one, where its exact value depends on the degree exponent $\lambda$
\cite{Schwartz-2002}. Since this value is clearly different from our experimental
observations and simulations, we conclude that the connectivity distribution in the
neural network is not a power law one. This may be a crucial difference between networks
grown with cultured neurons versus those growing naturally in the brain.

We thank L. Gruendlinger, M. Segal, J.-P. Eckmann, and O. Feinerman for their insight. J.
S. is supported by the Training Network PHYNECS, No. HPRN-CT-2002-00312. Work supported
by the Israel Science Foundation grant 993/05 and the Minerva Foundation, Germany.


\begin{thebibliography}{100}

\bibitem{Barabasi-2002} R. Albert and A.L. Barabasi, Rev. Mod. Phys. {\bf 74}, 47 (2002); S. H. Strogatz. Nature {\bf 410}, 268 (2001).

\bibitem{EckmannMoses-2002}
J. Eckmann and E. Moses. Proc. Natl. Acad. Sci. USA, {\bf 99},
5825 (2002); J.P. Eckmann, E. Moses, and D. Sergi. Proc Natl.
Acad. Sci. USA, {\bf 101}, 14333 (2004).

\bibitem{HybridProtein-2005}E.M. Marcotte  {\it et al.} Science {\bf 285}, 751 (1999).

\bibitem{White-1986} J.G. White {\it et al.} Phil. Trans. R. Soc. Lond. B {\bf 314}, 1 (1986).

\bibitem{Chan-2004} L. C. Jia {\it et al.} Phys. Rev. Lett. {\bf 93}, 088101 (2004).

\bibitem{Markram-2003} N. Kalisman {\it et al.} Biol, Cybern. {\bf 88}, 210 (2003).

\bibitem{Marom-2002} S. Marom and G. Shahaf. Q. Rev. Biophys. {\bf 35},63 (2002).

\bibitem{Jimbo-1999} Y. Jimbo {\it et al.}
Biophys. J. {\bf 76} 670 (1999).

\bibitem{Segev-2004} R. Segev {\it et al.} Phys. Rev. Lett. {\bf 92}, 118102 (2004).

\bibitem{Ofer-2006} O. Feinerman {\it et al.} J Neurophysiol. {\bf 94}, 3406
(2005); O. Feinerman and E. Moses, J. Neurosci. {\bf 26}, 4526 (2006).

\bibitem{Maeda-1995} E. Maeda {\it et al.} J. Neurosci. {\bf15}, 6834
(1995).

\bibitem{Stauffer} D. Stauffer and A. Aharony, Introduction to Percolation
Theory, 2nd ed. (Taylor \& Francis, London, 1991).

\bibitem{Newman-2001} M. E. J. Newman {\it et al.} Phys. Rev. E {\bf 64}, 026118 (2001); M. E. J. Newman. SIAM Review {\bf 45}, 167 (2003).

 \bibitem{Papa-1995} M. Papa {\it et al.} J. Neurosci. {\bf 15},
 1 (1995).

\bibitem{Supplementary} See EPAPS Document No. --- for details. For more information on EPAPS, see
http://www.aip.org/pubservs/epaps.html.

\bibitem{Shante-1971} F. Harary and G.E. Uhlenbeck. Proc. Nat.
Acad. Sci. {\bf 39} 315 (1952);
 V.K.S. Shante and S. Kirkpatrick. Adv. Phys.
{\bf 20}, 325 (1971).

\bibitem{Honore-1988} T. Honore {\it et al.} Science {\bf 241}, 701 (1988).


\bibitem{Schwartz-2002} N. Schwartz {\it et al.} Phys. Rev. E {\bf 66}, 015104(R) (2002).




\end{thebibliography}
\end{document}